\documentclass{ws-ijgmmp}

\begin{document}

\markboth{A. Odesskii, V. Rubtsov, V. Solkolov}
{Parameter-dependent AYBE and Poisson Brackets}

\def\C{\mathbb{C}}
\def\N{\mathbb{N}}
\def\Z{\mathbb{Z}}
\def\D{\mathbb{D}}
\def\R{\mathbb{R}}
\def\l{\lambda}
\def\g{\mathfrak g}
\def\vk{\varkappa}
\newcommand{\tcm}{\textcolor{magenta}}
\newcommand{\tcb}{\textcolor{blue}}
\def\dsize{\displaystyle}
\def\fs{\footnotesize}
\def\nn{ \nonumber }
\def\p{ \partial }
\def\defeq {\stackrel{\mbox{\rm\small def}}{=}}

\def\bq{ \begin{equation} }
\def\eq{ \end{equation} }
\def\ben{ \begin{eqnarray} }
\def\en{ \end{eqnarray} }

\def\frac#1#2{{#1\over #2}}
\def\dfrac#1#2{{\displaystyle{#1\over#2}}}
\def\on#1#2{\mathop{\vbox{\ialign{##\crcr\noalign{\kern2pt}
$\scriptstyle{#2}$\crcr\noalign{\kern2pt\nointerlineskip}
\kern-2pt$\hfil\displaystyle{#1}\hfil$\crcr}}}\limits}

\renewcommand{\theequation}{\arabic{section}.\arabic{equation}}
\newtheorem{prop}{Proposition}
\newtheorem{theor}{Theorem}
\newtheorem{Defi}{Definition}
\newtheorem{Lem}{Lemma}
\newtheorem{Res}{Result}
\newtheorem{Rem}{Remark}

\font\Sets=msbm10
\def\Complex {\hbox{\Sets C}}
\def\Real {\hbox{\Sets R}}
\def\Nat {\hbox{\Sets N}}
\def\Integ {\hbox{\Sets Z}}

\def\ldb{\mathopen{\{\!\!\{}} \def\rdb{\mathclose{\}\!\!\}}}
\def\ldbg{\mathopen{\bigl\{\!\!\bigl\{}} \def\rdbg{\mathclose{\bigr\}\!\!\bigr\}}}
\def\ldbgg{\mathopen{\Bigl\{\!\!\Bigl\{}} \def\rdbgg{\mathclose{\Bigr\}\!\!\Bigr\}}}

%
\catchline{}{}{}{}{}
%

\title{PARAMETER-DEPENDENT ASSOCIATIVE YANG-BAXTER EQUATIONS AND POISSON BRACKETS}

\author{ALEXANDER ODESSKII}

\address{Brock University,  Department of Mathematics,\\
St. Caterine, Ontario, Canada\\
aodesski@brocku.ca}

\author{VLADIMIR RUBTSOV}

\address{LAREMA, UMR 6093 du CNRS, D\'epartement de Math\'ematics\\
Universit\'e d'Angers, 2, bd. Lavoisier, 49045, Angers, Cedex 01, France\\
and\\
ITEP, 25, Bol. Cheremushkinskaya, 117259, Moscou, Russia\\
volodya@univ-angers.fr}

\author{VLADIMIR SOKOLOV}

\address{ Landau Institute for Theoretical Physics,\\
2, Kosygina, street, Moscow, Russia\\
vokolov@landau.ac.ru}
\maketitle

\begin{history}
\received{(Day Month Year)}
\revised{(Day Month Year)}
\end{history}

\begin{abstract}
We discuss associative analogues of classical Yang-Baxter equation meromorphically dependent on parameters. We discover that such equations 
enter in a description of a general class of parameter-dependent Poisson structures and double Lie and Poisson structures in sense of M. Van den Bergh.
We propose a classification of all solutions for one-dimensional associative Yang-Baxter equations. 
\end{abstract}

\keywords{Poisson brackets; double Poisson structures; Yang-Baxter equations.}

 \section{Introduction}

 Let ${\cal A}$ be an associative unital algebra over $\C$ containing a Lie algebra $\mathfrak g.$
 The parameter-dependent quantum $R-$ matrix $R(u)$  is an  invertible $\cal A\otimes \cal A-$valued solutions of Quantum Yang-Baxter Equation (QYBE). 
 This equation is  a functional relation depending on a complex parameter $u$ 
 $$R^{12}(u-v)R^{13}(u)R^{23}(v) = R^{23}(v)R^{13}(u)R^{12}(u-v) $$ which plays an important role in the theory
 of quantum integrable systems.  Roughly speaking, any quantum $R-$matrix gives rise to a quantum integrable system.

 A classical analogue of the QYBE - the Classical Yang-Baxter Equation (CYBE) can be conventionally written
 as 
 $$[r_{12}(u-v),r_{13}(u)+r_{23}(v)]+[r_{13}(u),r_{23}(v)] =0,$$
 where $r(u)$ is in the following relation with $R(u)$:
 $$R(u) = \mathbb I\otimes \mathbb I + \hbar r(u) + O(\hbar).$$
It is clear that $r(u)$  is a $\mathfrak g\otimes \mathfrak g-$valued function.

The problem of finding solutions to QYBE and CYBE (quantum  $R-$matrices and classical $r-$matrices)  
was one of central and intriguing problems in modern Mathematical Physics  during the last forty years.  There are many beautiful and explicit results describing the solutions of the YBE's. 
The first general classification of solutions to the CYBE for complex simple Lie algebras is due to Belavin and Drinfeld (\cite{BD}).  They showed that there exist three different types of the parameter dependence
for the classical $r-$matrix:  rational, trigonometric and elliptic.

The Quantum Inverse Scattering Method provides the existence of intimate relations between the YBE and Hamiltonian (Poisson) structures of the corresponding integrable systems.
For example, the classical $r-$matrix induces a Poisson bracket on the corresponding Lie group $G={\rm Lie}(\mathfrak g)$ such that the group operation is a Poisson morphism 
(a Poisson-Lie group structure of V. Drinfeld).

A third version of  YBE: a (parameter-dependent)  Associative Yang-Baxter Equation (AYBE) (see the definition below) had appeared independently  in attempts to understand a 
nature of the associativity constraint or "Massey triple relations" in the framework of the homology mirror symmetry on elliptic curves (1999-2000,\cite{Polis}) and (as a constant or 
parameter independent) counterpart of an AYBE - in a description of some associative version of the Lie algebra Drinfeld double (2000, \cite{Agui}).
We remark also that the similar constant AYBE emerges in the paper of  S. Fomin and  A. Kirillov . They have introduced a model for the cohomology ring of the flag manifold $X={\rm SL}_n/B$, using a non-commutative algebra  defined by the quadratic relations satisfied by the divided difference operators corresponding to the positive roots.  (1995-96 \cite{AK}).  One of the relations in their algebra coincides with those in  (\cite{Agui}). 
We should note also that  a 1-dimensional multiplicative AYBE analogue of  (\cite{Polis}) was discovered by B. Feigin and one of the authors (A.O.) as  the exchange relations 
for elliptic Sklyanin algebras (1993, \cite{OdFeig1}).

An interesting relations of CYBE and AYBE with vector bundle geometry on singular and degenerated  cubic curves were described in papers of  I. Burban with collaborators (\cite{Burban}). 
They extend the ideas and results of A. Polishschuk to much wider class of plane cubic curves and their degenerations. They find in this frame many solutions of  the matrix-valued AYBEs and CYBEs  
appeared within their algebro-geometric picture. 

One should remark that the solutions of  AYBE equations sometimes are related to QYBE and CYBE but are essentially different. Solving the AYBE we solve often in the same time a CYBE and QYBE but not vice-versa.

As we have observed above a motivation to study various YBE comes from the  Integrable Systems. 
A natural question of an associative algebra analogue for the  Hamiltonian formalism and the related Poisson structures arises in the theory of integrable systems with matrix variables.
An intuitive version of it was defined by  one of the authors (V.S.) with various collaborators in (\cite{MikSok} , \cite{OlSok}).  More extended answer was developed in  (\cite{ORS1}) and can be reduced to a study of Poisson structures on the representations of the initial associative algebra, following the Kontsevich "Representation functor" philosophy. (\cite{Konts}).  The important role in this reduction belongs to M. Van den Bergh and his 
\emph{double Lie and Poisson structures.} (see \cite{VdB}) Any such double Poisson structure defines a Lie algebra structure on the trace space $\cal A/[\cal A, \cal A]$ of the initial algebra $\cal A$. We have studied (and classified for  "low-dimensional" case ) such {\emph quadratic} (double) Poisson structures in the case of the free associative algebra,  clarified their connection with the constant AYBE and studied the corresponding \emph{trace-brackets}  in (\cite{ORS}). In this context our results are complementary to those in (\cite{Sched}).

The goal of this paper is to discuss generalisations of the above concepts and relations between them to the parameter-dependent case.
We discuss  the parameter-dependent AYBE which appears in different setups of mathematical physics. We describe the parameter-dependent Poisson and double Poisson brackets and we establish the connexions
between them and the (parameter-dependent) AYBE.  It turns out that linear and quadratic brackets are closely related to a general associative algebra analogue of the AYBE introduced by A. Polishchuk in (\cite{Polis})
in the case of the matrix algebra. This observation allows us  to construct examples of parameter-dependent double brackets using solutions of AYBE found in (\cite{OdSok}). Some of these solutions are generating functions for families of constant double and trace Poisson brackets. The direct relation between the parameter-dependent associative and classical $r-$ matrices and Poisson brackets seems quite natural and 
it is one of the main results of the paper. We could not find in the literature any trace of the relations between the parameter-dependent 
$r-$matrices which solve the corresponding AYBE and some parameter-dependent Poisson structures as well as a connection with double Lie and Poisson structures.

We propose also a full classification of  the parameter-dependent AYBE solutions for the case of $m=1.$ Some of our solutions can be easily generalised to higher dimension cases and coincide with solutions obtained by A. Polishchuk and by I. Burban with co-authors.

The volume limit does not permit us to consider many other aspects of the parameter-dependent Poisson and double Poisson structures and their connections with functional associative algebras. In particular, the bi-hamiltonian properties and constraints  and the corresponding examples of their solutions are out of the scope of the paper. We hope to come back to these questions in other publications.
\section{Associative Yang-Baxter equation} 
\subsection{Constant AYBE}
By definition, a classical $r$-matrix on an associative $\C-$ algebra ${\cal A}$ is any solution of the following tensor equations 
\begin{equation}\label{tens1c}
r^{12}=-r^{21}
\end{equation}
and
\begin{equation}\label{tens12}
r^{12}  r^{23} = r^{13} r^{12} +
 r^{23}r^{13}.
\end{equation}
Here  $r \in  {\cal A} \otimes {\cal A}$ and (\ref{tens2}) is a relation defined in the tensor cube ${\cal A} \otimes {\cal A}  \otimes {\cal A}$. Relation 
(\ref{tens12}) is called {\it a classical associative Yang-Baxter equation.}
 
Rewrite (\ref{tens12}) as $A(r):= r^{12}  r^{23} - r^{13} r^{12} - r^{23}r^{13}=0$, apply the "flip"  operator $P^{13}$ (the transposition on the first and third factors)
to (\ref{tens12}) and use the skew-symmetry (\ref{tens1}). We obtain the "conjugated" AYBE $A^*(r) =  r^{23}  r^{12} - r^{13} r^{23} - r^{12}r^{13}=0.$
The difference $[[r]] := A(r)-A^*(r)$ is a skew-symmetric element in $\mathfrak g\otimes \mathfrak g\otimes \mathfrak g.$ The equation
\begin{equation}\label{CYBE}
[[r]]:=[r^{12},r^{13}] + [r^{12}r^{23}] + [r^{13}r^{23}] =0.
\end{equation}
is nothing but the {\it constant classical Yang-Baxter equation.}

In the case when $\cal A= {\rm Mat_m}(\C)$ the equations (\ref{tens1}), (\ref{tens12}) are equivalent to 
\begin{equation}\label{r2}
r^{\lambda\sigma}_{\alpha\beta}
r^{\mu\nu}_{\sigma\tau}+r^{\mu\sigma}_{\beta\tau} r^{\nu\lambda}_{\sigma\alpha}+r^{\nu\sigma}_{\tau\alpha} r^{\lambda\mu}_{\sigma\beta}=0,
\qquad r^{\sigma \epsilon}_{\alpha\beta}=-r^{\epsilon\sigma}_{\beta\alpha}.
\end{equation}
Here $r=r^{km}_{ij} e^{i}_{k} \otimes e^{j}_{m}$, where $e^{i}_{j}$ are the matrix unities:
$e^j_i e^m_k=\delta^j_k e^m_i.$ 

The tensor $r$ may be also interpreted 
 
 1) as an operator on $\C^m\otimes \C^m$; 
 
 2) as an operator on ${\rm Mat}_m(\C)$.

For the first interpretation all operators act in $\C^m \otimes \C^m \otimes \C^m$,  $\sigma^{ij}$ 
means the transposition of $i$-th and $j$-th components of the tensor
product, and $r^{ij}$ means the operator $r$ acting in the
product of the $i$-th and $j$-th components.  

For the second interpretation, we define operators $r, \bar r : Mat_{m}(\C)\to Mat_{m}(\C)$ by 
 $\quad r(x)^p_q=r^{m p}_{n q} x^{n}_{m}$, \quad $\bar r(x)^p_q=r^{p m}_{n q} x^{n}_{m}.$ 
 Then (\ref{dub1}), (\ref{dub2}) provide the following operator identities:  
$$ 
\begin{array}{c}
 r(x)=-{\bar r}(x), \qquad r(x) r(y)=r(x r(y))+r(x) y),\\[2mm]
 \qquad \bar r(x) \bar r(y)=\bar r(x \bar r(y))+\bar r(x) y),\\[2mm]
 \end{array}
$$ for any $x,y$.  These identities mean that  operators  $r$ and $\bar r$ satisfies the {\it Rota-Baxter equation} (\cite{Rota}) and this  fact implies also
that the new matrix multiplications $\circ_r$ and $\circ_{\bar r}$ defined by
\begin{equation} \label{mult}
x\circ_{r}y= r(x)y + xr(y),\qquad x\circ_{\bar r}y= {\bar r}(x)y + x{\bar r}(y)
\end{equation}
are associative.

\subsection{General AYBE with four parameters}

The most interesting applications of classical $r$-matrices related to Lie algebras are refered to the parameter-dependent case. 
In general, $r$ depends on two parameters $u,v$ but usually one assumes that $r=r(u-v).$ Remarkably, the most general associative $r$-matrix depends on four parameters!  

{\bf Definition.} (cf. the matrix algebra case in \cite{Polis} and \cite{Burban}) 
A classical parameter dependent $r$-matrix on an associative algebra ${\cal A}$ is any solution of the following functional equations 
{\begin{equation}\label{tens1}
r^{12}(u,v,x,y)=-r^{21}(v,u,y,x)
\end{equation}
and
\begin{equation}\label{tens2}
r^{12}(u,v,x,y) r^{23}(u,w,y,z) = r^{13}(u,w,x,z) r^{12}(w,v,x,y) +
 r^{23}(v,w,y,z) r^{13}(u,v,x,z).
\end{equation}
Here  $r : \C^2\times \C^2  \to {\cal A} \otimes {\cal A}$ is a meromorphic $\cal A \otimes \cal A-$ valued function and (\ref{tens2}) is a relation defined in the tensor cube ${\cal A} \otimes {\cal A}  \otimes {\cal A}$. 

This definition is the straightforward generalisation of the four parameter-dependent unitary AYBE which was defined in (\cite{Polis}) and extensively studied in (\cite{Burban}). Let us remind the context of the
A. Polischuk's definition.
 Let $\cal E$ be a smooth (for simplicity) projective elliptic curve and $E\in {\rm Vect}(\cal E)$ will be always a simple (${\rm End}(E) = \C$) vector bundle on it. The fibre of $E$ in a point $x\in \cal E$ will be denote by
$E_x$. Let $E_1,E_2\in Vect(\cal E)$ be two  simple vector bundles such that ${\rm Ext(E_1, E_2) = 0}.$  Then for distinct points $x_1,x_2 \in \cal E$ one can define  a tensor
\begin{equation}\label{ellr-mat}
r_{x_1,x_2}^{E_1,E_2}\in {\rm Hom}(E_{1,x_1}^*, E_{2,x_1}^*)\otimes {\rm Hom}(E_{2,x_2}^*, E_{1,x_2}^*)
\end{equation}
such that for a triple of simple vector bundles $E_1,E_2,E_3 \in {\rm Vect}(\cal E)$ pairwise satisfying the above conditions and for a triple of distinct points $x_1,x_2,x_3\in \cal E$ the
following relations  in ${\rm Hom}(E_{1,x_1}^*, E_{2,x_1}^*)\otimes {\rm Hom}(E_{2,x_2}^*, E_{3,x_2}^*)\otimes {\rm Hom}(E_{3,x_3}^*, E_{1,x_3}^*)$ satisfies:
\begin{equation}\label{ell-unit}
(r_{x_1,x_2}^{E_1,E_2})^{21} = - r_{x_2,x_1}^{E_2,E_1}
\end{equation}
and
\begin{equation}\label{ell-AYBE}
(r_{x_1,x_2}^{E_3,E_2})^{12}(r_{x_1,x_3}^{E_1,E_3})^{13} - (r_{x_2,x_3}^{E_1,E_3})^{23}(r_{x_1,x_2}^{E_1,E_2})^{12} + (r_{x_1,x_3}^{E_1,E_2})^{13}(r_{x_2,x_3}^{E_2,E_3})^{23}=0
\end{equation}
(see \cite{Polis}.)
This tensor expresses a certain  triple Massey product in the derived category of $\cal E$ and the relation (\ref{ell-AYBE}) is a consequence of the associativity constraint corresponding to the $A_{\infty}-$ structure
related to this category. 
To get the parameter-dependent AYBE one should uniformize the curve $\cal E$, trivialise all vector bundles and express the tensor (\ref{ellr-mat}) as a  ${\rm Mat}_m (\C)\otimes {\rm Mat}_m (\C)-$ valued 
function of complex variables $r_{12}(u,v,x,y)$ where $(u,v)$ correspond to $E_1, E_2$ and $(x,y)-$ to $x_1,x_2$ (see \cite{Polis}).
It should be stressed that the AYBE (\ref{ell-AYBE}) gives the AYBE depending on the differences of the variables and the $r-$matrix $r_{12}(u,v,x,y) = r_{12}(u-v,x-y).$

\section{Quadratic algebras depending on parameters} 
\subsection{Quadratic algebra with two parameters.}
The following equation has appeared  at the first time in the paper  of B. Feigin and one of the authors (A.O.) (\cite{OdFeig1}), where a quadratic algebra 
with generators ${\bf e}(u,x)$ and commutation relations
$$
{\bf e}(u,x) {\bf e}(v,y)=\beta(u,v,x,y)  {\bf e}(v,y)  {\bf e}(u,x)+\alpha(u,v,x,y)  {\bf e}(u,y)  {\bf e}(v,x)
$$
has been considered. 
A priori it was assumed that the coefficients $\alpha$ and $\beta$ only depend on the differences $u-v$ and $x-y$ . It is possible to verify that the coefficients $\alpha$ and $\beta$ have the following structure:
$$
\beta(u,v,x,y)=\frac{\psi(u,v)}{\phi(x,y)},  \qquad \alpha(u,v,x,y)=\frac{r(u,v,x,y)}{\phi(x,y)},
$$
where $r$ satisfies the conditions
\bq\label{rC1} r(u,v,x,y)= - r(v,u,y,x)\eq
and
\bq\label{rC2}
r(w,u,z,x) r(v,w,y,x)+  
r(u,v,x,y) r(w,u,z,y)+  
r(v,w,y,z) r(u,v,x,z)=0,
\eq
which coincide (for $\cal A=\C$) with (\ref{tens1}), (\ref{tens2}). Solutions of equations (\ref{rC1}), (\ref{rC2})  are discussed in Section 4.
The functions $\phi$ and $\psi$ are determined by additional equations 
$$
r(u,v,x,y) r(u,v,y,x)=\phi(x,y) \phi(y,x)-\psi(u,v) \psi(v,u).
$$
and
$$
(\psi(v,w) \psi(w,v)-\psi(u,v) \psi(v,u))\, r(u,w,x,z)= 
$$
$$
r(u,v,x,y) r(u,v,y,z) r(v,w,x,z)-
r(u,v,x,z) r(v,w,x,y) r(v,w,y,z).
$$
It turns out that  such functions $\psi$ and $\phi$ exist for any solution of (\ref{rC2}),(\ref{rC1}).

\subsection{Quadratic Poisson brackets.}
Our first observation is  that  the {\it same} associative Yang-Baxter appears in the classical version of the above "quantum" construction. 
Consider the intertwining operation of the form 
\bq\label{quadPoiss}\{e_i (u,x), e_j (v,y) \}= \beta_{ij}^{ml}(u,v,x,y) e_m(u,x)e_l(v,y)+r_{ij}^{ml}(u,v,x,y) e_m(u,y)e_l(v,x),\eq
where $i,j=1,...,m.$

\begin{theorem}
The operation for (\ref{quadPoiss}) satisfies all axioms of a Poisson bracket iff 
\bq\label{secondPoiss0} 
 r_{ij}^{ml}(u,v,x,y)= - r_{ji}^{lm}(v,u,y,x), 
\eq
and
$$\beta_{ij}^{ml}(u,v,x,y)= - \beta_{ji}^{lm}(v,u,y,x).$$
and
\bq\label{secondPoiss1} 
r^{\sigma q}_{ki}(w,u,z,x) r^{r s}_{j \sigma}(v,w,y,x)+  
r^{\sigma r}_{ij}(u,v,x,y) r^{s q}_{k \sigma}(w,u,z,y)+  
r^{\sigma s}_{jk}(v,w,y,z) r^{q r}_{i \sigma}(u,v,x,z)=0,
\eq
with
\bq\begin{array}{c}\label{secondPoiss2}
r^{\sigma q}_{ij}(u,v,x,y) \beta^{r s}_{k \sigma}(w,u,z,y)-  
r^{\sigma s}_{ji}(v,u,y,x) \beta^{r q}_{k \sigma}(w,v,z,x) \,+
\\[3mm]
\beta^{\sigma r}_{jk}(v,w,y,z) r^{s q}_{i \sigma}(u,v,x,y)-  
\beta^{\sigma r}_{ik}(u,w,x,z) r^{q s}_{j \sigma}(v,u,y,x)  
=0,
\end{array}
\eq 
and with 
\bq\begin{array}{c}\label{secondPoiss3}
\beta^{\sigma q}_{jk}(v,w,y,z) \beta^{r s}_{i \sigma}(u,v,x,z) -  
\beta^{\sigma r}_{ji}(v,u,z,x) \beta^{q s}_{k \sigma}(w,v,z,y) \,+
\\[3mm]
\beta^{\sigma r}_{ki}(w,u,z,x) \beta^{s q}_{j \sigma}(v,w,y,x)-  
\beta^{\sigma s}_{kj}(w,v,x,y) \beta^{r q}_{i \sigma}(u,w,x,z)\,+ 
\\[3mm]  
\beta^{\sigma s}_{ij}(u,v,x,y) \beta^{q r}_{k \sigma}(w,u,z,y)-  
\beta^{\sigma q}_{ik}(u,w,y,z) \beta^{s r}_{j \sigma}(v,u,y,x)
=0.
\end{array}
\eq
are satisfied.
\end{theorem}

If we denote
$$r(u,v,x,y)=r^{jm}_{ik}(u,v,x,y) e^i_{j}\otimes e^k_{m}, $$ where $e^i_{j}$ are matrix unities in ${\rm Mat}_m (\C)$, 
then equations (\ref{secondPoiss1}) are nothing but the associative Yang-Baxter equation (\ref{tens2}) for $\cal A={\rm Mat}_m (\C).$

The equation (\ref{secondPoiss3}) is a  parameter-dependent analogue of the classical Yang-Baxter equation  (CYBE) (\ref{CYBE})  which is a difference of two AYBE:
$$[[\beta,\beta]] = A(\beta) - A^*(\beta),$$
where $\beta(u,v,x,y) =  \beta_{ij}^{ml}(u,v,x,y) e^i_{m}\otimes e^j_{l}$  and $A(\beta)$ is exactly the "positive sign" part of the equation (\ref{secondPoiss3}) while  $A^*(\beta)$ is its "conjugation" which coincides with the "negative sign" part of the same equation.

In the case  $m=1$ this equation coincides with (\ref{rC2}), the equation (\ref{secondPoiss3}) are trivially satisfied and the equation (\ref{secondPoiss2}) is reduced to
$$
\beta(w,u,z,y)-\beta(w,u,z,x)+\beta(v,w,y,z)-\beta(v,w,x,z)=0.
$$ 
The general skew-symmetric solution of this equation is given by
$$
\beta(u,v,x,y)=a(u,v)+b(x,y)+c(v,x)-c(u,y),
$$
where $a(u,v)=-a(v,u),$ $b(x,y)=-b(y,x).$

\subsection{Double Poisson brackets with parameters}

Our second observation is that relations (\ref{secondPoiss0}), (\ref{secondPoiss1}) describe a class of double Poisson brackets (see (\cite{VdB}) with two parameters.  
Let us remind the basic definition.

\begin{Defi}\label{doublbr}  (M. Van den Bergh). A double Poisson bracket on an associative algebra ${\cal A}$ is a $\C$-linear map
$\ldb,\rdb : {\cal A} \otimes {\cal A} \mapsto  {\cal A} \otimes {\cal A}$ satisfying the following conditions:
\bq \label{dub1}
\ldb u, v\rdb = - \ldb v,u\rdb ^{\circ},
\eq
\bq \label{dub2}
 \ldb u, \ldb v,w \rdb \rdb_l + \sigma  \ldb v, \ldb w,u \rdb \rdb_l +\sigma^2  \ldb w, \ldb u,v \rdb \rdb_l  =0,
\eq
and
\bq\label{dub3}
\ldb u, vw \rdb = (v\otimes 1)  \ldb u,w \rdb  + \ldb u,v \rdb (1\otimes  w).
\eq
\end{Defi}
Here $(u\otimes v)^{\circ}:=v\otimes u$; $\ldb v_1, v_2\otimes v_3 \rdb_l :=\ldb v_1, v_2 \rdb \otimes v_3$
 and $\sigma( v_1 \otimes v_2 \otimes v_3 ):= v_{3}\otimes v_{1} \otimes v_{2}$.

If the bracket $\ldb,\rdb : {\cal A} \otimes {\cal A} \mapsto  {\cal A} \otimes {\cal A}$ satisfies only to (\ref{dub1}) and (\ref{dub2}) one can  say that this is just a {\em double Lie bracket}.

The main property of double Poisson bracket for us is the following relation between double and usual Poisson brackets established by M.Van den Bergh  (\cite{VdB}).  
Let $\mu$ denote the multiplication map $\mu:  {\cal A}\otimes {\cal A} \to {\cal A}$ i.e. $\mu(u\otimes v)=u v.$
We define a $\C-$bilinear bracket operation in ${\cal A}$ by  $\{-,-\}:=\mu(\ldb -, - \rdb).$

\begin{prop} Let $\ldb -, - \rdb$ be a double Poisson bracket on ${\cal A}$. Then $\{-,-\}$ is an $H-$ Poisson bracket of W. Crawley-Boevey on $\cal A/[\cal A, \cal A]$ (\cite{CB}) which is defined as
\bq\label{trbr}
\lbrace \bar a, \bar b\rbrace =\overline{\mu(\ldb a,b\rdb},
\eq
where $\bar a$ means the image of $a\in {\cal A}$ under the natural projection ${\cal A}\to {\cal A}/[{\cal A},{\cal A}].$
\end{prop}

\begin{Rem}
$H-$ Poisson bracket of W. Crawley-Boevey (strictly speaking) is not a Poisson bracket (the quotient of "traces" $\cal A/[\cal A, \cal A]$ is not an algebra) but it satisfies the following important
property: for any $b\in A$ the endomorphism $[{\bar b}, ]$ of $\cal A/[\cal A, \cal A]$ is induced by some derivative $\partial_b$ of $\cal A:$
$$[{\bar b}, \bar c] = \overline {\partial_b (c)}, \quad c\in \cal A.$$
\end{Rem}
Consider  a double bracket operation defined on generators ${\bf E}(u,x),$ where $u,x\in \C,$ by the formula 
 \bq\label{doublegener}\begin{array}{c}
\ldb {\bf E}_i(u,x), {\bf E}_j(v,y) \rdb = \\[4mm] 
\alpha^{km}_{ij}(u,v,x,y) {\bf E}_k(u,y)\otimes {\bf E}_m(v,x) + \beta^{km}_{ij}(u,v,x,y) {\bf E}_k(v,x)\otimes {\bf E}_m(u,y)+
\\[4mm] 
 \gamma^{km}_{ij}(u,v,x,y) {\bf E}_k(u,x)\otimes {\bf E}_m(v,y) + \delta^{km}_{ij}(u,v,x,y) {\bf E}_k(v,y)\otimes {\bf E}_m(u,x).
\end{array}
\eq

\begin{theorem}
The operation (\ref{doublegener}) gives an example of (parameter-dependent) double Lie algebra iff
$$
\alpha_{ij}^{km}(u,v,x,y)=-\alpha_{ji}^{mk}(v,u,y,x), \qquad \beta_{ij}^{km}(u,v,x,y)=-\beta_{ji}^{mk}(v,u,y,x),
$$
$$
\gamma_{ij}^{km}(u,v,x,y)=-\gamma_{ji}^{mk}(v,u,y,x), \qquad \delta_{ij}^{km}(u,v,x,y)=-\delta_{ji}^{mk}(v,u,y,x)
$$
and the following chain of relations satisfies:
\bq\label{alph-alph}\begin{array}{c}
\alpha_{jk}^{\sigma m}(v,w,y,z)\alpha_{i\sigma}^{pq}(u,v,x,z) + \alpha_{ki}^{\sigma p}(w,u,z,x)\alpha_{j\sigma}^{qm}(v,w,y,x)+
\\[4mm] 
+ \alpha_{ij}^{\sigma q}(u,v,x,y)\alpha_{k\sigma}^{mp}(w,u,z,y) =0;
\end{array}
\eq
\bq\label{bet-bet}\begin{array}{c}
\beta_{jk}^{\sigma m}(v,w,y,z)\beta_{i\sigma}^{pq}(u,w,x,y) + \beta_{ki}^{\sigma p}(w,u,z,x)\beta_{j\sigma}^{qm}(v,u,y,z) +
\\[4mm]
+ \beta_{ij}^{\sigma q}(u,v,x,y)\beta_{k\sigma}^{mp}(w,v,z,x) =0;
\end{array}
\eq
\bq\label{alph-bet}
\alpha_{jk}^{\sigma m}(v,w,y,z)\beta_{i\sigma}^{pq}(u,v,x,z) =0;
\eq
\bq\label{bet-alph}
\beta_{jk}^{\sigma m}(v,w,y,z)\alpha_{i\sigma}^{pq}(u,w,x,y) =0;
\eq
\bq\label{del-del}\begin{array}{c}
\delta_{jk}^{\sigma m}(v,w,y,z)\delta_{i\sigma}^{pq}(u,w,x,z)  + \delta_{ki}^{\sigma p}(w,u,z,x)\delta_{j\sigma}^{qm}(v,u,y,x)+
\\[4mm] 
+ \delta_{ij}^{\sigma q}(u,v,x,y)\delta_{k\sigma}^{mp}(w,v,z,y) =0;
\end{array}
\eq
\bq\label{gam-gam}\begin{array}{c}
\gamma_{jk}^{\sigma m}(v,w,y,z)\gamma_{i\sigma}^{pq}(u,v,x,y) + \gamma_{ki}^{\sigma p}(w,u,z,x)\gamma_{j\sigma}^{qm}(v,w,y,z) +
\\[4mm]
+ \gamma_{ij}^{\sigma q}(u,v,x,y)\gamma_{k\sigma}^{mp}(w,u,z,x) =0;
\end{array}
\eq
\bq\label{alph-gam}
\alpha_{jk}^{\sigma m}(v,w,y,z)\gamma_{i\sigma}^{pq}(u,v,x,z) + \gamma_{ki}^{\sigma p}(w,u,z,x)\alpha_{j\sigma}^{qm}(v,w,y,z) =0;
\eq
\bq\label{bet-gam}
\beta_{jk}^{\sigma m}(v,w,y,z)\gamma_{i\sigma}^{pq}(u,w,x,y) + \gamma_{ki}^{\sigma p}(w,u,z,x)\beta_{j\sigma}^{qm}(v,w,y,z) =0;
\eq
\bq\label{alph-del}
\alpha_{jk}^{\sigma m}(v,w,y,z)\delta_{i\sigma}^{pq}(u,v,x,z) + \delta_{ij}^{\sigma q}(u,v,x,y)\alpha_{k\sigma}^{mp}(w,v,z,y) =0;
\eq
\bq\label{bet-del}
\beta_{jk}^{\sigma m}(v,w,y,z)\delta_{i\sigma}^{pq}(u,w,x,y) + \delta_{ij}^{\sigma q}(u,v,x,y)\beta_{k\sigma}^{mp}(w,v,z,y) =0;
\eq
\bq\label{gam-del}
\gamma_{jk}^{\sigma m}(v,w,y,z)\delta_{i\sigma}^{pq}(u,v,x,y) + \delta_{ij}^{\sigma q}(u,v,x,y)\gamma_{k\sigma}^{mp}(w,v,z,y) =0.
\eq
\end{theorem}
The first four relations ( a skew-symmetry) are resulted  from (\ref{dub1}) and the Jacobi identity (\ref{dub2}) implies the other  functional relations

\begin{Rem}
If  $\gamma = \delta =0$ the double brackets (\ref{doublegener})  are called the (purely) exchange brackets.
\end{Rem}
\begin{Rem}
If  we extend this double Lie structure by the Leibniz rule of  van den Bergh (\ref{dub3}) we obtain a parameter dependent double Poisson structure on the functional algebra $\cal A.$
In what follows we shall always suppose such extension when we speak about various examples of double Poisson structures.
\end{Rem}

\subsubsection{Quadratic one-parameter double brackets}

Consider the quadratic double Poisson brackets in the case when the generators depend on only one parameter. More general then  (\ref{doublegener}) skew-symmetric ansatz for such bracket has the form 
\begin{equation}\begin{array}{c}
\ldb {\bf E}_i(u), {\bf E}_j(v) \rdb = \alpha^{pq}_{ij}(u,v) {\bf E}_p(u)\otimes {\bf E}_q(v)+  \beta^{pq}_{ij}(u,v) {\bf E}_p(v)\otimes {\bf E}_q(u)+
\\[4mm] 
\gamma^{pq}_{ij}(u,v) {\bf E}_p(u) {\bf E}_q(v)\otimes 1+  \delta^{pq}_{ij}(u,v) 1 \otimes {\bf E}_p(u) {\bf E}_q(v)-
\\ [4mm]
\delta^{pq}_{ji}(v,u) {\bf E}_p(v) {\bf E}_q(u)\otimes 1-  \gamma^{pq}_{ji}(v,u) 1 \otimes {\bf E}_p(v) {\bf E}_q(u),
\end{array} \label{double}
\end{equation}
where 
$$
\alpha^{pq}_{ki}(u,v)=- \alpha^{qp}_{ik}(v,u), \qquad \beta^{pq}_{ki}(u,v)=- \beta^{qp}_{ik}(v,u).
$$
If the tensors $\gamma$ and $\delta$ vanish then (\ref{double}) looks like a special case of (\ref{doublegener}). However 
the conditions for (\ref{double}) to be a double Poisson bracket is weaker then for (\ref{doublegener}). Indeed,
it follows from  (\ref{dub2}) that $\alpha$ and $\beta$ should satisfy equations 
\begin{equation}\begin{array}{c} \label{albet}
\alpha^{\sigma q}_{ki}(w,u) \alpha^{r s}_{j \sigma}(v,w)+  
\alpha^{\sigma r}_{ij}(u,v) \alpha^{s q}_{k \sigma}(w,u)+  
\alpha^{\sigma s}_{jk}(v,w) \alpha^{q r}_{i \sigma}(u,v)=0; \\[4mm] 
\beta^{\sigma q}_{ki}(w,u) \beta^{r s}_{j \sigma}(v,u)+  
\beta^{\sigma r}_{ij}(u,v) \beta^{s q}_{k \sigma}(w,v)+  
\beta^{\sigma s}_{jk}(v,w) \beta^{q r}_{i \sigma}(u,w)=0; \\[4mm] 
\alpha^{\sigma q}_{jk}(v,w) \beta^{r s}_{i \sigma}(u,v)+ 
\beta^{\sigma s}_{ij}(u,v) \alpha^{q r}_{k \sigma}(w,v)=0.
 \end{array}
\end{equation}
The third of these conditions differs from (\ref{alph-bet}) and (\ref{bet-alph}).

It is easy to see that first equation (\ref{albet}) describes solutions of (\ref{tens2}) that do not depend on two first arguments whereas 
solutions of the second equation (\ref{albet}) correspond to solutions of (\ref{tens2}) that do not depend on third and fourth arguments.  
Such solutions of (\ref{tens2}) have not been systematically studied.

The other relations between the coefficients in (\ref{double}) read as follows:
$$
\begin{array}{c}
\delta^{pq}_{kj}(w,v) \delta^{rs}_{ip}(u,w) =\delta^{rp}_{ik}(u,w) \delta^{sq}_{pj}(w,v);  \\[4mm] 
\gamma^{pq}_{jk}(v,w) \gamma^{rs}_{pi}(v,u)=0, \qquad \gamma^{qp}_{ki}(w,u) \gamma^{rs}_{jp}(v,u)=0; \\[4mm] 
\gamma^{qp}_{ij}(u,v) \gamma^{rs}_{kp}(v,w)=0, \qquad \gamma^{pq}_{jk}(v,w) \gamma^{rs}_{ip}(u,v)=0; \\[4mm] 
\delta^{qp}_{ji}(v,u) \delta^{rs}_{kp}(w,u)+\delta^{rs}_{pi}(w,u) \beta^{pq}_{jk}(v,w)+\delta^{ps}_{ji}(v,u) \beta^{qr}_{kp}(w,v)=0; \\[4mm] 
 \delta^{pq}_{ji}(v,u) \delta^{rs}_{pk}(v,w) +\delta^{rs}_{jp}(v,w) \alpha^{pq}_{ki}(w,u)+ \delta^{rp}_{ji}(v,u) \alpha^{qs}_{pk}(u,w)=0; \\[4mm] 
 \end{array}
$$ 
$$
\gamma^{rs}_{ip}(u,v) \alpha^{pq}_{jk}(v,w)+\gamma^{ps}_{ij}(u,v) \alpha^{qr}_{kp}(w,u)=\gamma^{rp}_{ki}(w,u) \beta^{sq}_{jp}(v,u) + \gamma^{rs}_{pi}(w,u) \beta^{qp}_{kj}(w,v)=0;
$$
$$
\delta^{pq}_{ik}(u,w) \alpha^{rs}_{jp}(v,u) +\delta^{sq}_{pk}(u,w) \alpha^{pr}_{ij}(u,v)=\delta^{rs}_{ip}(u,w) \beta^{pq}_{jk}(v,w) + \delta^{rp}_{ik}(u,w) \beta^{qs}_{pj}(w,v)=0;
$$
$$
\gamma^{qr}_{pk}(v,w) \delta^{ps}_{ji}(v,u) - \gamma^{qp}_{jk}(v,w) \delta^{rs}_{pi}(w,u)=\gamma^{rs}_{ip}(u,v) \delta^{qp}_{kj}(w,v) - \gamma^{pr}_{ij}(u,v) \delta^{qr}_{kp}(v,w)=0;
$$
$$
\gamma^{qp}_{ij}(u,v) \alpha^{sr}_{pk}(v,w)  + \gamma^{ps}_{ij}(u,v) \delta^{qr}_{pk}(u,v)=\gamma^{rs}_{pj}(w,v) \alpha^{pq}_{ki}(w,u)  + \gamma^{rs}_{kp}(w,v) \delta^{pq}_{ji}(v,u)=0;
$$
$$
\gamma^{pq}_{ij}(u,v) \beta^{rs}_{kp}(w,u) + \gamma^{qp}_{ij}(u,v) \delta^{rs}_{kp}(w,v)=\gamma^{rs}_{ip}(u,w) \beta^{pq}_{jk}(v,w) + \gamma^{rs}_{pk}(u,w) \delta^{qp}_{ji}(v,u)=0.
$$

In the case $m=1$ we get $\gamma(u,v)=0,$ and 
\bq\label{doub2}\begin{array}{c}
\alpha(v,w) \alpha(u,v)+\alpha(w,u) \alpha(v,w)+\alpha(u,v) \alpha(w,u)=0, \\[4mm] 
\beta(v,w) \beta(u,v)+\beta(w,u) \beta(v,w)+\beta(u,v) \beta(w,u)=0, \\[4mm] 
\delta(v,u) \delta(v,w)=\alpha(w,u) \Big(\delta(v,u)-\delta(v,w)\Big), \\[4mm]
\delta(w,v) \delta(u,v)=\beta(w,u) \Big(\delta(w,v)-\delta(u,v)\Big). \\[4mm]
\end{array}
\eq
The general solution is given by
\begin{equation}\label{gh}
\alpha(u,v)=\frac{1}{g(u)-g(v)}, \quad \beta(u,v)=\frac{1}{h(u)-h(v)}, \quad \delta(u,v)=\frac{\epsilon}{g(v)-h(u)}.
\end{equation}
Here $g$ and $h$ are arbitrary functions of one variable and the constant $\epsilon$ equals 1 or 0. 

In the case of arbitrary $m$ there exist particular solutions of the form 
$$
\alpha^{ij}_{pq}(u,v)=\frac{k^{ij}_{pq}}{u-v}, \qquad \beta^{ij}_{pq}(u,v)=-\alpha^{ij}_{pq}(u,v), \qquad \gamma^{ij}_{pq}(u,v)=\delta^{ij}_{pq}(u,v)=0.
$$
where 
\begin{equation}\label{case11}
k^{ij}_{pq}=\delta^i_p \delta^j_q
\end{equation}
or
\begin{equation}\label{case21}
k^{ij}_{pq}=\delta^i_q \delta^j_p. 
\end{equation}
The corresponding double brackets are given by
$$
\ldb {\bf E}_i(u), {\bf E}_j(v) \rdb=\frac{{\bf E}_i(u)\otimes {\bf E}_j(v)-{\bf E}_i(v)\otimes {\bf E}_j(u) }{u-v}
$$
and
$$
\ldb {\bf E}_i(u), {\bf E}_j(v) \rdb=\frac{{\bf E}_j(u)\otimes {\bf E}_i(v)-{\bf E}_j(v)\otimes {\bf E}_i(u) }{u-v}.
$$

Solutions (\ref{case11}), (\ref{case21}) are analogs of the Yang solutions in the theory of classical Yang-Baxter equation on Lie algebras (\cite{reysem}). It is known that in the Lie case the sum of the Yang solution and any constant solution is a solution of the same equation.
In our situation it is easy to prove the following

{\bf Proposition 2}\,\,(cf. \cite{OdSok}). Let $r$ be any solution of (\ref{r2}). Then 
\begin{equation}\label{case1}
\alpha^{ij}_{pq}(u,v)=\frac{\delta^i_p \delta^j_q}{u-v}, \qquad \beta^{ij}_{pq}(u,v)=\frac{\delta^i_p \delta^j_q}{v-u}+r^{ij}_{pq}
\end{equation}
and
\begin{equation}\label{case2}
\alpha^{ij}_{pq}(u,v)=\frac{\delta^i_q \delta^j_p}{u-v}+r^{ij}_{pq}, \qquad \beta^{ij}_{pq}(u,v)=\frac{\delta^i_q \delta^j_p}{v-u}
\end{equation}
satisfy (\ref{albet}). $\square$

Formulas (\ref{case1}) and (\ref{case2}) generate families of constant trace Poisson brackets (\cite{ORS1}). To get them one should set ${\bf E}_i(x)=\sum_0^n e_{ij}x^j$ and apply the construction of Proposition 1. As a result, some trace brackets on free associative algebra with $m n$ generators $e_{ij}$ will be defined.

\section{Solutions in the case $m=1$ and their classification.}

If $m=1$  the following solutions of functional equations  (\ref{secondPoiss0}), (\ref{secondPoiss1})
 
 \bq \label{sol1} \displaystyle r(u,v,x,y) = \frac{1}{u-v} - \frac{1}{x-y}; \eq
 \bq \label{sol2} \displaystyle r(u,v,x,y) = \frac{e^{u-v} - e^{x-y}}{(e^{u-v} -1)(e^{x-y} -1)}; \eq
  \bq \label{sol3} \displaystyle r(u,v,x,y) =\frac{\theta_{11}(u-v+x-y)}{\theta_{11}(u-v)\theta_{11}(x-y)}; \eq 
 where
 $$\theta_{11}(u) = \sum_{n\in \mathbb Z}(-1)^n \exp{(\pi i(n+\frac{1}{2})^2\tau + 2\pi i (n+\frac{1}{2})u)},$$
have been found in \cite{OdFeig1} and independently in  \cite{Polis}. Here $\theta_{11}(u)$ is the quasi-periodic Jacobi theta-function on the lattice $\Z \oplus {\tau}\Z$ and $\Im \tau>0$.
Notice that all these solutions have the form $r=r(u-v,x-y). $
For arbitrary $m$ some solutions of the form $r=r(u-v,x,y)$ have been obtained in \cite{Burban} by pure geometrical methods. 
Deep general statements concerning the relations of such solutions with quantum ${\mathfrak{gl}}(m)$ $r$-matrices, $A_{\infty}$-constraints and mirror conjecture on elliptic 
curves have been discovered in \cite{Polis}. However no relations of these solutions with two-parametric Poisson and double Poisson brackets was not observed.  

We present a classification of solutions in the case $m=1$ up to 
the following equivalence transformations
 \begin{equation}\label{equiv1}
r(u,v,x,y)\to r(u,v,x,y)\, \frac{q(u,y) q(v,x)}{q(u,x) q(v,y)}
 \end{equation}
 and 
  \begin{equation}\label{equiv2}
u\to \phi(u), \qquad v\to \phi(v), \qquad x\to \psi(x), \qquad y\to \psi(y)\
 \end{equation}
where $q,$ $\phi,$ $\psi$ are arbitrary functions.

We claim the following
\begin{theorem}
Any solution $r(u,v,x,y)$ of (\ref{secondPoiss0}),(\ref{secondPoiss1}) that has simple poles at $u=v$ and
$x=y$ is equivalent to one of solutions (\ref{sol1})-(\ref{sol3}).
\end{theorem}
We split the proof in four lemmas:
 
 {\bf Lemma 1.} Any solution of (\ref{rC1}), (\ref{rC2}) can be reduced by a transformation of the form (\ref{equiv1}) to a solution of  the following form
 \begin{equation}\label{separ}
 r(u,v,x,y)=f(u,v)+g(x,y)+h(u,x)-h(v,y),
 \end{equation}
 where $f,g,h$ are some functions of two variables such that $f(u,v)=-f(v,u)$, $g(x,y)=-g(y,x)$. $\square$ 
 
 Assume that  both $f$ and $g$ have a simple pole at $u=v$ and $x=y$ correspondingly. Substituting (\ref{separ}) into (\ref{rC2}) and expanding the obtained expressions at $y=x$ and at $v=u$, we prove the following statements. 
 
 {\bf Lemma 2.} If $f$ and $g$ have a simple pole at $u=v$ and $x=y$ then up to a transformation (\ref{equiv2})
 \begin{equation}\label{separ1}
 f(u,v)=\frac{ U(u)+ U(v)}{u-v}, \qquad g(x,y)=\frac{ V(x)+ V(y)}{x-y},
 \end{equation}
 where 
 \begin{equation}\label{separ2}
 U(x)=\sqrt{P}, \quad P(x)= \sum_{i=0}^4 k_i x^i, \qquad V(x)=\sqrt{Q}, \quad Q(x)= \sum_{i=0}^4 r_i x^i. \qquad \square
 \end{equation}
 
 Consider the case $P\ne 0, \, Q\ne 0.$ 
 
 {\bf Lemma 3.} The function $h$ has the form
  \begin{equation}\label{separ3}
 h(u,x)=A_1(u)-A_2(x)+Z(B_1(u)-B_2(x)),
 \end{equation}
 where
 $$
 A_1'=\frac{- V'^2-V V''}{6\, V},\qquad A_2'=\frac{- U'^2-U U''}{6\, U},  \qquad B_1'=\frac{1}{V}, \qquad B_2'=\frac{1}{U},
 $$
 and
  \begin{equation}\label{ZZ}
 Z''^2=2 Z'^3+\lambda_1 Z'+\lambda_2
 \end{equation}
for some constants $\lambda_i,$ depending on the coefficients of $P$ and $Q$.   $\square$

 {\bf Lemma 4.} For the polynomials $P$ and $Q$ the functions
 $$
 I_1(P)=k_2^2-3 k_1 k_3+12 k_0 k_4, \qquad I_2(P)=2 k_2^3-9 k_1 k_2 k_3+27 k_0 k_3^2+27 k_1^2 k_4-72 k_0 k_2 k_4
 $$
 have the same values and 
  \begin{equation}\label{lam}
\lambda_1= -\frac{I_1}{6}, \qquad \lambda_2=\frac{I_2}{108}. \qquad \square
 \end{equation}
 
 Lemmas 2-4 describe all solutions of the form (\ref{separ}) under assumption  $P\ne 0, \, Q\ne 0.$ To bring these solutions to a canonical form we may use fraction-linear transformations of the form
 \begin{equation}\label{frac}
 u\to \frac{a_1 u+b_1}{c_1 u+d_1}, \quad  v\to \frac{a_1 v+b_1}{c_1 v+d_1}, \qquad x\to \frac{a_2 x+b_2}{c_2 x+d_2}, \quad  y\to \frac{a_2 y+b_2}{c_2 y+d_2}.
 \end{equation}
It is easy to verify that $I_1$ and $I_2$ are semi-invariants with respect to (\ref{frac}).  It is clear that 
$P$ can be reduced by (\ref{frac}) to one of the following canonical forms:  $P(x)=1,\,$  $P(x)=x,\,$  $P(x)=x^2,\,$  $P(x)=x (x-1),\,$  and $P(x)=x (x-1) (x-\sigma).$ 

Moreover, the canonical form $P=x (x-1)$ can be reduced to $P=x^2$ by the transformation $x \to -\frac{(2 x-1)^2}{8 x}$ while $P=x$ is related to $P=1$ by  $x \to \frac{x^2}{4}.$ Thus we may consider only three canonical forms for $P$: $P=1$, $P=x^2$ and  $P=x (x-1) (x-\sigma)$ and the same canonical forms for $Q$.  Taking into account Lemma 4, we can verify that these canonical forms can be combined in a unique way: 1) $P=Q=1$; 2) $P=Q=x^2$ and  3) $P=Q=x (x-1) (x-\sigma)$.

The solutions corresponding to the cases 1) and 2) can be written (up to the equivalence) in the following
 uniform way: 
 $$r(u,v,x,y) = \frac{p_1 u v+p_2 (u+v)+p_3}{(u-v)}- \frac{p_1 x y+p_2 (x+y)+p_3}{(x-y)}.$$ 
Any such solution is equivalent to (\ref{sol1}) or to (\ref{sol2}).  

 
 In the case $P(x)=x (x-1) (x-\sigma)$ we perform the transformation  $u\to \wp(u),v\to \wp(v),x\to \wp(x),y\to \wp(y)$ and write  the solution in the form 
$$ 
r= \frac{1}{\theta_{11}'(0}\left(\frac{\theta_{11}'(u-v)}{\theta_{11}(u-v)} -\frac{\theta_{11}'(u+x+\eta)}{\theta_{11}(u+x+\eta)}+
 \frac{\theta_{11}'(x-y)}{\theta_{11}(x-y)} - \frac{\theta_{11}'(v+y+\eta)}{\theta_{11}(v+y+\eta)}\right),
$$
where $\eta$ is arbitrary parameter. This solution can rewritten as 
$$
r  =\frac{\theta_{11}(u-v+x-y)\theta_{11}(u+y+\eta)\theta_{11}(v+x+\eta)}{\theta_{11}(u-v)\theta_{11}(x-y)\theta_{11}(u+x+\eta)\theta_{11}(v+y+\eta)},
$$
which is equivalent to (\ref{sol3}). 

It can be verified that in the case $U\ne 0,$ $V=0$ any solution is equivalent to $r(u,v,x,y)=\frac{1}{u-v}$ while
if $U=0,$ $V\ne0$ we arrive at   $r(u,v,x,y)=\frac{1}{x-y}.$

\section{Linear parameter-dependent Poisson brackets}

\subsection{One parameter-dependent Poisson brackets}

We remind the conditions for one-parameter Poisson algebra with generators ${\bf E}_i(u)$, where $i=1,...,m$ and $u\in \C:$
$$\lbrace {\bf E}_i(u), {\bf E}_j(v) \rbrace = \alpha_{ij}^{ml}(u,v){\bf E}_m(u){\bf E}_l(v),$$
where
$$\alpha_{ij}^{ml}(u,v) = -\alpha_{ji}^{lm}(v,u)$$
from the skew-symmetry constraint.
Then the Jacobi identity gives the following 6-term relation:
\bq\begin{array}{c}\label{oneparamPoiss}
\alpha_{ij}^{\sigma l}(u,v)\alpha_{k\sigma}^{rs}(w,u)+\alpha_{jk}^{\sigma r}(v,w)\alpha_{i\sigma}^{sl}(u,v)+\alpha_{ki}^{\sigma s}(w,u)\alpha_{j\sigma}^{lr}(v,w) -\\[3mm]
-\alpha_{ji}^{\sigma s}(v,u)\alpha_{k\sigma}^{rl}(w,v)-\alpha_{kj}^{\sigma l}(w,v)\alpha_{i\sigma}^{sr}(u,w)-\alpha_{ik}^{\sigma r}(u,w)\alpha_{j\sigma}^{ls}(v,u)=0.
\end{array}
\eq
The equation (\ref{oneparamPoiss}) is a full analogue of the classical Yang-Baxter equation  (CYBE) which is a "difference" of two AYBE:
$$[[\alpha,\alpha]] = A(\alpha) - A^*(\alpha),$$
where $A(r)$ is exactly the LHS of the first equation from (\ref{albet}) while  $A^*(r)$ is its "conjugation" which coincides with the LHS of the second equation
in (\ref{albet}) if we replace $\beta$ by $\alpha$.

In the one-parameter case our free associative algebra $\cal A$ is generated by ${\bf E}_i(u)$, where $i=1,...,m$ and $u\in \C$ is a continues parameter. Any local double bracket is defined by brackets between the generators. The locality means that
$$
\ldb {\bf E}_i(u), {\bf E}_j(v) \rdb = s^{km}_{ij}(u,v) {\bf P}_k\otimes {\bf P}_m
$$
where $P_i$ are some (non-commutative) polynomials in  ${\bf E}_1(u),{\bf E}_1(v),...,{\bf E}_m(u),{\bf E}_m(v).$ As usual it the theory of parameter dependent Poisson brackets, the coefficients $s^{km}_{ij}(u,v)$ may have singularities at $u=v.$

\subsection{Linear parameter-dependent double brackets and associativity conditions.}

The general ansatz for local linear double Poisson brackets is given by  
\begin{equation}\begin{array}{c}
\ldb {\bf E}_i(u), {\bf E}_j(v) \rdb = a_{ij}^k(u,v) {\bf E}_k(u)\otimes 1 +  b_{ij}^k (u,v){\bf E}_k(v)\otimes 1 +
\\[4mm] 
c_{ij}^k (u,v)1\otimes {\bf E}_k(u)  + d_{ij}^k (u,v)1 \otimes {\bf E}_k(v)
\end{array} \label{linear}
\end{equation}It follows from property (\ref{dub1}) that 
$c_{ij}^k(u,v)=-b_{ji}^k(v,u), d_{ij}^k(u,v)=-a_{ji}^k(v,u).$
The (twisted) Jacobi identity (\ref{dub2}) is equivalent to  
the following functional relations for the coefficients:
\begin{equation} \label{albetlin} \begin{array}{c}
a_{ij}^{\sigma}(u,v)a_{k\sigma}^{\lambda}(w,u) -  a_{ki}^{\sigma}(w,u)a_{\sigma j}^{\lambda}(w,v) + a_{k\sigma}^{\lambda} (w,v)b_{ij}^{\sigma}(u,v) =0;\\[4mm] 
b_{ij}^{\sigma}(u,v)b_{k\sigma}^{\lambda}(w,v)    - b_{ki}^{\sigma}(w,u)b_{\sigma j}^{\lambda}(u,v) - b_{\sigma j}^{\lambda}(w,v) a_{ki}^{\sigma} (w,u) = 0;\\[4mm]
a_{jk}^{\sigma} (v,w)b_{i\sigma}^{\lambda}(u,v)  - b_{ij}^{\sigma}(u,v)a_{\sigma k}^{\lambda}(v,w)=0.\\[4mm] 
\end{array} 
\end{equation}
\begin{Rem} The same conditions are equivalent to the associativity of the following product:
$$
{\bf E}_i(u) {\bf E}_j(v)=a_{ij}^k(u,v) {\bf E}_k(u)+b_{ij}^k(u,v) {\bf E}_k(v). 
$$
\end{Rem}
\qquad In the case $m=1$ we get the following two functional equations to $a$ and $b$:
$$\begin{array}{c}
a(u,v)a(w,u) -  a(w,u)a(w,v) + a (w,v)b(u,v) =0;\\[4mm]  
b(u,v)b(w,v)    - b(w,u)b (u,v) - b(w,v) a (w,u) = 0.
\end{array}
$$
If $a b\ne 0,$ the general answer for these equations is of the form 
$$
a(u,v)=\frac{a_1(v)b_1(u)}{b_1(u)-b_1(v)}, \qquad  b(u,v)=\frac{a_1(u)b_1(v)}{b_1(v)-b_1(u)},
$$
where $a_1,b_1$ are arbitrary functions of one variable. 

The simplest solution of (\ref{albetlin}) for arbitrary $m$ is given by
\begin{equation}\label{simsol}
a_{ij}^{k}(u,v)=\frac{c_{ij}^{k}}{u-v}, \qquad b_{ij}^{k}(u,v)=\frac{c_{ij}^{k}}{v-u},
\end{equation}
where $c_{ij}^{k}$ are structural constants of any $m$-dimensional associative algebra ${\cal A}$.

 More general, let $r(u,v): {\cal A}\to {\cal A}$ be any solution of the parameter dependent Rota-Baxter equation
\begin{equation} \label{ybass1}
(r(u,w)x)(r(u,v)y)-r(u,v)((r(v,w)x)y)-r(u,w)(x(r(w,v)y))=0
\end{equation}
introduced in (\cite{OdSok}). Then
$$
a_{ij}^k(u,v)=c_{is}^k r_{j}^s(u,v), \qquad b_{ij}^k(u,v)=c_{sj}^k r_{i}^s(v,u)
$$
satisfy (\ref{albetlin}). Some solutions of equation (\ref{ybass1}) have been found in (\cite{OdSok}).

In the case $\cal A={\rm Mat}_m (\mathbb C)$ the Rota-Baxter equation (\ref{ybass1}) has the form 
\bq\label{odsok}
r_{\beta j}^{\alpha p}(u,v) r^{r \sigma}_{s p}(u,w)=r^{r p}_{s \beta}(v,w) r^{\alpha \sigma}_{p j}(u,v)+ r^{p \sigma}_{s j}(u,w) r^{\alpha r}_{\beta p}(w,v).
\eq
This equation is a partial case of (\ref{tens2}) under the additional skew-symmetry assumption - (\ref{tens1}):  
$$r^{jm}_{ik}(u,v)=-r^{mj}_{ki}(v,u).$$.}
\begin{Rem} The parameter dependent Rota-Baxter equation (\ref{ybass1}) is an associativity condition of  the second multiplication operation
$$x\circ_{r(u,v)} y := r(u,v)(x)y +xr(u,v)(y)$$ in the algebra ${\cal A}$ which becomes {\it doubly associative algebra} in terminology  of
M. Semenov-Tyan-Shansky (who had proposed the first example of  generalized parameter-dependent Rota-Baxter equation in  (\cite{reysem}).
His example deals with the associative algebra ${\cal A}={\rm Mat}_n\otimes \C[u,u^{-1}]$ of matrices over the Laurent polynomial ring.
His operator $r(u,v)$ is defined  as identity on the subalgebra of matrix with non-negative degrees in $u$ and minus identity on the subalgebra of matrix with strictly negative
degrees in $u$. This operator satisfies  the special case of general Rota-Baxter weight $\lambda$ relation, (here $\lambda$ is a complex parameter):
\begin{equation} \label{ybassgen}
(r(u,w)x)(r(u,v)y)-r(u,v)((r(v,w)x)y)-r(u,w)(x(r(w,v)y)) = \lambda xy.
\end{equation}
The equation (\ref{ybass1}) corresponds the case $\lambda = 0$ and the Semenov-Tian-Shansky Rota-Baxter operator corresponds to $\lambda = -1.$
\footnote{We are thankful to M. Semenov-Tian-Shansky who had drawn our attention to the fact of the first appear of the parameter-dependent doubly associative
algebra in (\cite{reysem}) and its classical roots.}
\end{Rem}

{\bf Acknowledgments.}
The authors would like to thank   I. Burban, M. Kontsevich, A. Polischchuk, T. Schedler and M. Semenov-{T}ian-{S}hansky
for useful discussions. 
The results of the paper became a subject of an invited lecture of V.R. during XXII Workshop in Geometry and Physics in
september 2013 in \'Evora (Portugal). He is grateful to the organisers for this invitation.
 VS and VR  are acknowledged the  hospitality of MPIM(Bonn) where the paper was started. They are grateful to {\rm Mat}PYL 
project "Non-commutative integrable systems" and the ANR ``DIADEMS'' project for a financial support of VS visits in Angers. 
They were also  partially supported by the RFBR grant 11-01-00341-a. V.R. thanks to the grant FASI RF 14.740.11.0347 and RFBR grant
12-01-00525. AO and VS are thankful to IHES for its support and hospitality.

\end{document}